\documentstyle[12pt]{article}
\def\beq{\begin{equation}}
\def\eeq{\end{equation}}

\begin{document}

\begin{titlepage}
\begin{flushright}
SNUTP--94--49  \\  YUMS-94--17 \\ (May 1994)  \\

\end{flushright}
\vspace{0.4in}
\begin{center}
{\Large \bf  New Approach  for  Measuring $| V_{ub} |$ \\  at Future
$B$--Factories \\}
\vspace{0.8in}
{\bf  C.~S.~ Kim$^{a}$,~~~Pyungwon Ko$^{b}$
\\
Daesung Hwang$^{a}$ and Wuk Namgung$^{c}$ \\ }
\vspace{0.4in}
{\sl
$^{a}$ Department of Physics, Yonsei Univeristy, Seoul 120-749, KOREA  \\
$^{b}$ Department of Physics, Hong-Ik University, Seoul 121-791, KOREA \\
$^{c}$ Department of Physics, Dongkook Univerity, Seoul, KOREA   \\
}
\vspace{0.4in}
{\bf   Abstract  \\ }

\end{center}
\noindent
It is suggested that the measurements of hadronic invariant mass  ($m_X$)
distributons in the inclusive $B \rightarrow X_{c(u)} l \nu$ decays can
be useful in extracting the CKM matrix element $|V_{ub}|$.
We investigated hadronic invariant mass distributions within  the various
theoretical models of  HQET, FAC and chiral lagrangian as well as ACCMM model.
It is also emphasized that the $m_X$ distribution even at the region
$m_{X} > m_{D}$ in the inclusive $b\rightarrow u$ are effetive in selecting the
events, experimentally viable at the future asymmetric $B$ factories,
with better theoretical understandings.

\end{titlepage}

\noindent
{\Large  \bf 1. Introduction \\}

The CKM matrix element $V_{ub}$ is important to the standard model
description  of CP violation.
If it were zero, there would be no CP violation from the CKM matrix
element ({\it i.e.} in the standard model), and we have to seek for other
source of CP violation in $K_{L} \rightarrow \pi\pi$.
Observations of semileptonic $b\rightarrow u$ transitions by the CLEO
\cite{cleo1} and ARGUS \cite{argus1}  imply that $V_{ub}$ is  indeed nonzero,
and it is important to
extract the modulus $|V_{ub}|$ from semileptonic decays of $B$ mesons
as accurately as possible.    Presently, the charged lepton  energy is
measured, and the $b\rightarrow u$ events are selected from the high
end of the charged lepton energy spectrum.  This method is applied to both
inclusive and exclusive semileptonic $B$ decays.

However, this cut on $E_l$ is not very effective, since only  below $ 10 \%$
of $b\rightarrow u$   events survive this cut.
As first discussed in \cite{cskimetal},   the measurements of
hadronic invariant mass ($m_X$) distributions in $B \rightarrow X_{c,u} l
\nu$ (inclusive decays)  can be useful to extract a CKM matrix element
$V_{ub}$.  For $b \rightarrow c l \nu$, one necessarily has
$m_{X} \geq m_{D} = 1.86$ GeV.  So, if we impose a condition $m_{X}
< m_{D}$, the resulting events come from $b \rightarrow u l \nu$.
According to the work of \cite{cskimetal},  $\sim$ 90\% of the $b\rightarrow
u$ events survive this cut.   This is in sharp contrast with the usual
cut on $E_l$.  Thus, one could get an order of magnitude improvement in
selecting data.

Inclusive $m_X$ distributions for $b\rightarrow u$ transition
can be obtained using the ACCMM model \cite{accmm}, or the work by BUSV
\cite{busv}.
In this work, we follow the result of Ref.~\cite{cskimetal}
based on the   ACCMM model,  
with a remark that
theoretical uncertainties for the inclusive $b\rightarrow u$ decays are
less compared to an   exclusive mode.

Resonance contributions to the $m_X$ distributions in  $B \rightarrow
X_{c,u} l \nu$ can be estimated  invoking  various models.
Once the decay rate for $B \rightarrow R l \nu$ (where $R$ is a resonance)
is known,  the corresponding $m_X$  distribution can be written as
\begin{equation}
{d \Gamma \over d m_X} \approx {2 m_{X} \Gamma (B\rightarrow R l \nu) \over
\pi}~{m_{R} \Gamma_{R} \over
(m_{X}^{2} - m_{R}^{2})^{2} + m_{R}^{2} \Gamma_{R}^{2} },
\end{equation}
in the narrow width approximation.  Here, $m_{R}$ and $\Gamma_R$ are the
mass and the width of the resonance $R$. In the limit of $ \Gamma_{R}
\rightarrow 0$, we get
\begin{equation}
{d \Gamma \over d m_X} = \Gamma (B\rightarrow R)~\delta (m_{X} - m_{R}),
\end{equation}
so that  the correct decay rate for $B \rightarrow R l \nu$ comes out upon
integrating Eq.~(2)   over $dm_X$.

In Section 2, we discuss
the result for $B\rightarrow D, D^{*}, (D^{**})$  in the heavy quark
effective theory.  In Section 3,  $B\rightarrow (\pi,\rho) l \nu$ are
considered in two different types of approaches: a nonrelativistic  quark
model and the chiral lagrangian with heavy mesons as well as light vector
mesons.  In Section 4, the $m_X$ distributions for
$B\rightarrow X_{c (u)} l \nu$ are
shown, and it is emphasized that the $m_X$ distribution even at the region
$m_{X} > m_{D}$ in the inclusive $b\rightarrow u$ are effetive in selecting
almost $ 100 \%$  of  the
events, experimentally viable at the future asymmetric $B$ factories.
Since one can calculate the inclusive decay more reliably,  one can
achieve better determination of $V_{ub}$ both statistically and
systematically.

\vspace{.5in}

\noindent
{ \Large \bf 2. The $B\rightarrow X_{c} l \nu$ decay in the heavy quark
effective  theory \\}

Let us first consider $B \rightarrow X_{c} l \nu$, which is   dominated by
resonance contributions with $X_{c} = D, D^{*}, D^{**}$.
Theoretical predictions based on the heavy
quark effective field theory (HQET) \cite{neubert} depend on one hadronic
form factor   $h_{A_1} (w)$.
In order to calculate a decay rate, one has to know the shape of the form
factor over the whole kinematic range  of $w$.
However, this form factor is not calculable from the first principle,
except for the normalization : $h_{A_1} (1) = 0.99 \pm 0.04$, which is one
of the predictions  of HQET \cite{neubert}.
Thus, one necessarily resort to some model for the shape of the form factor.
If one adopts the result of the QCD sum rule results, one may approximate
the form factor as
\begin{equation}
h_{A_1} (w) \approx 0.99 ~\left( { 2 \over  w + 1 } \right)^{ 2
\varrho_{A_1}^2},
\end{equation}
with $ \varrho_{A_1}^2  \approx 0.8$.
Since this result is based on the QCD sum rule, there exist some
systematic uncertainties associated with the  sum rule.
This systematic uncertainty may be taken into account by allowing
$ \varrho_{A_1}^2 $ vary between  0.5 and 1.1,  where the  latter
is on the border of the limit given by Voloshin's sum rule \cite{voloshin}.
For this range of $ \varrho_{A_1}^2$, one can predict the decay rates
for  $B \rightarrow D^{(*)} l \nu$ \cite{neubert} :
\begin{eqnarray}
\Gamma (B\rightarrow D l \nu) & = & (0.86 \sim 1.35) \times 10^{13} ~
|V_{bc}|^{2} / {\rm sec}, \\
\Gamma (B\rightarrow D^{*} l \nu) & = & (2.59 \sim 3.48) \times 10^{13} ~
|V_{bc}|^{2} / {\rm sec},
\end{eqnarray}
where the smaller decay rates correspond to the larger $\varrho_{A_1}^2$.
For $B \rightarrow D^{**} l \nu$,  we use the observation by CLEO
\cite{cleoold}:
\begin{equation}
Br ( B \rightarrow D^{**} l \nu ) \simeq 0.5~Br ( B \rightarrow D l \nu).
\end{equation}
Our  approach  concerning  the measurement of $|V_{ub}|$  from the $m_X$
distributions can be regarded independently of  the uncertainties in Eqs.
(4)--(6),
because the $m_X$ distributions from the three resonances $D, D^*$ and$D^{**}$
are essentially delta functions,  Eq. (2),  and we just require to
exclude small regions  around $m_{X} = m_{D}, m_{D^{*}}, m_{D^{**}}$.
For more details, see Section 4 and Fig. 1.

\vspace{.5in}

{\Large  \bf 3. The $B\rightarrow X_{u} l \nu$ decay \\}

Unlike the $b\rightarrow c$ transition, the $b\rightarrow u$ transition
is largely nonresonant and multiple jet-like final states dominate
\cite{donoghue1}.
The whole inclusive decay can be theoretically well understood in most
of the kiematical region.  The electron energy spectrum  or the hadronic
mass distribution for the inclusive semileptonic decay can be calculated
rather reliably.  In contrast, for the exclusive decays for $b \rightarrow
u$ such as $B \rightarrow (\pi, \rho) + l \nu$,  the model dependence
becomes more pronounced, especially for the shape of the form factors.
Here, we consider two classes of models, the FAC model (a nonrelativistic
quark model) and the chiral lagrangian with heavy mesons.  The results
are compared with the $m_X$  distribution obtained by the ACCMM model
in Section 4.

\vspace{.3in}

\noindent
{\bf A. The FAC model \\  }

The FAC model is based on the nonrelativistic quark model with assuming the
form factors are  factorized as \cite{hagiwara}
\begin{equation}
f_{i}^{\rm FAC} (q^{2}) = f_{i}^{\rm FQM} (q^{2}) \times F(q^{2}),
\end{equation}
where $f_{i}^{\rm FQM} (q^{2})$ is the free quark model form factors
arising from the overlap of the spin wavefunctions, and $F(q^{2})$
comes from the overlap of the spatial wave functions.
For $B \rightarrow D^{(*)}$ transitions, one can impose the normalization
condition, $F(q_{max}^{2}) = 1$,
by heavy quark flavor symmetry. For $B\rightarrow \pi ({\rm or  }~ \rho)$
transitions,  this normalization is not valid and it may be reasonable to
assume that the form factor suppression for $B\rightarrow \pi$ begins at
the $B\rightarrow \rho$ threshold.
With these assumptions, one gets
\begin{eqnarray}
\Gamma (B \rightarrow D l \nu ) & = & (0.71 \sim 0.85) \times
10^{13} ~|V_{bc}|^{2} / {\rm sec},
\\
\Gamma (B \rightarrow D^{*} l \nu ) & = & (2.17 \sim 2.44) \times
10^{13} ~|V_{bc}|^{2}/ {\rm sec},
\\
\Gamma (B^{0} \rightarrow \pi^{+} l \nu ) & = & (0.24 \sim 0.86) \times
10^{13} ~|V_{ub}|^{2}/ {\rm sec},
\\
\Gamma (B^{0} \rightarrow \rho^{+} l \nu ) & = & (0.77 \sim 2.10) \times
10^{13} ~|V_{ub}|^{2}/ {\rm sec},
\end{eqnarray}
for certain ranges of pole masses (see Ref.~\cite{hagiwara} for more
details.)
Note that the FAC model predictions for $B\rightarrow D^{*}$ are consistent
with (although they are systematically lower than)  those by the HQET
discussed in the previous section.  Hence, the FAC model for the $B
\rightarrow D^{(*)}$ transitions are rather reliable.
For $B\rightarrow \pi ({\rm or } ~\rho)$ transitions, the predictions
are very sensitive to the shape of the form factors  because of the large
phase space available.  Therefore, we simply regard the above numbers for
$B\rightarrow \pi ({\rm or }~ \rho)$ transitions as exemplary values  in
a nonrelativistic quark model, without giving much meanings to the specific
values.

\vspace{.3in}

\noindent
{\bf B. The chiral lagrangian with heavy mesons \\}

Recently, the chiral lagrangian with heavy mesons and baryons has been
developed \cite{wise}--\cite{yan}.  This lagrangian was originally invented
in order to  describes interactions among heavy mesons and light mesons
such as $\pi$ and $K$ in the soft pion limit. Then, heavy baryons \cite{cho}
as well as  $\rho$ \cite{casalbuoni}--\cite{ko}
have been incorporated in the leading order in $1/m_Q$ and chiral expansions.
The weak current can be represented in terms of physical fields like heavy
hadrons and light mesons, allowing us to calculate the matrix element of
the weak current between hadrons and thus the semileptonic decays of heavy
hadrons.

However, this apporach has its own limitations.  First of all, the hadronic
form factors given by this chiral lagrangian is valid  only in very limited
regions of the whole kinematic region.
Therefore, one often assumes certain shape of form factors and normalize
them at a point to a value given by the chiral lagrangian with heavy hadrons.
Furthermore, if one considers the next--to--leading order corrections
in $1/m_Q$ and chiral expansions, there come in a lot of unknown parameters
and one essentially loose predictability.  Although the reparametrization
invariance of the heavy quark field leads to some constraints  to the
parameters in the next--to--leading order terms,  it still leaves many
other parameters unconstrained.  Therefore,  results based on the
chiral lagrangian with heavy baryons should be understood, keeping in
mind the uncertainties just mentioned above.

One of the extensive studies of semileptonic decays of heavy mesons in the
framework of the chiral lagrangian with heavy hadrons is the work by
R. Casalbuoni and his collaborators \cite{casalbuoni}.  Their results are
\begin{eqnarray}
\Gamma ( B^{0} \rightarrow \pi^{-} l \nu ) & = & 38.8 ~\left( f_{B}
({\rm MeV}) \over 200 \right)^{2} ~\times 10^{13} |V_{ub}|^{2} /{\rm sec},
\label{eq:casalpi}
\\
\Gamma ( B^{0} \rightarrow \rho^{-} l \nu ) & = & 22.7 ~\left( f_{B}
({\rm MeV}) \over 200 \right)^{2} ~\times 10^{13} |V_{ub}|^{2} /{\rm sec}.
\label{eq:casalrho}
\end{eqnarray}
At this point, a remark on the  $B-$meson decay constant  $f_B$ in
Eqs.~(\ref{eq:casalpi}) and (\ref{eq:casalrho})  is in order.
In the lowest order in the $1/m_Q$ expansion,
\begin{equation}
{f_{B} \over f_D} = \sqrt{ M_{D} \over M_B}.
\label{eq:scaling}
\end{equation}
On the other hand, the lattice QCD and the QCD sum rule \cite{lattice}
suggest that
\begin{equation}
f_{B} \approx f_{D} \approx 200~~{\rm MeV},
\label{eq:lattice}
\end{equation}
which violates the scaling relation, Eq.~(\ref{eq:scaling}).
Thus,  the results in Ref.~\cite{casalbuoni} are expressed as above,
although   it is not systematic in $1/m_Q$ expansion to use  Eq.~
(\ref{eq:lattice}).

We note that the results of Ref.~\cite{casalbuoni} are subtantially larger
than those based on the FAC model.  Especially, relative  ratios between
$B\rightarrow \pi$ and $B\rightarrow \rho$ are opposite in two models, and
may be checked in the near future.  For the isospin--related decay
$B^{-} \rightarrow \rho^{0} l^{-} \bar{\nu}_{l}$,   the predicted decay rate
is the half of  Eq.~(\ref{eq:casalrho}), with the corresponding branching
ratio
\begin{equation}
Br ( B^{-} \rightarrow \rho^{0} l^{-} \bar{\nu}_{l} ) = 0.44 \times 10^{-3}~
\left( f_{B} ({\rm MeV}) \over 200 \right)^{2} ~\left|{V_{ub} \over 0.0045}
\right|^2,
\end{equation}
assuming $\tau_{B} = 1.29$ ps.  The data from ARGUS and CLEO seem
contradictory with each other :
\begin{eqnarray}
Br (B^{-} \rightarrow \rho^{0} l^{-} \bar{\nu}_{l} )
& = & (1.13 \pm 0.36 \pm 0.26) \times 10^{-3}~~~~({\rm ARGUS})
\\
& < & 0.3 \times 10^{-3}~~~~~~~~~~~~~~~~~~~~~~~~~~({\rm CLEO}).
\end{eqnarray}
Note that two data are incompatible with each other.
The ARGUS result \cite{argus2} is consistent with the prediction by R.
Casalbuoni {\it et al.}, but is inconsistent with the FAC model prediction.
On the other hand, if the result by CLEO \cite{cleo2}  is confirmed,
the prediction based on the chiral lagrangian with heavy mesons would be
excluded.   In this case,  there can be many possible reasons for it.
First of all, interactions between  $\rho$ and heavy mesons may not be
well described by the chiral lagrangian in the lowest order because of
relatively heaviness of $\rho$.   This would be contrary to the better known
case,  the chiral lagrangian with vector mesons ($\rho$), where  dynamics
of $\pi,\rho$ are rather well described.  Secondly, and most likely,
the usual simple assumption on the shape of the form factor may not be
right.   In most cases including Ref.~\cite{casalbuoni},  it is assumed that
a form factor $f(q^{2})$ satisfies a monopole form :
\begin{equation}
f (q^{2}) = {f (0) \over 1 - q^{2} /M_{pole}^2 },
\end{equation}
where $M_{pole}$ is a pole mass.  This extrapolation of the form factor
through the whole kinematic range does not have justifications from the
first principle, and is a source of uncertainties in any models.

\vspace{.5in}

\noindent
{ \Large \bf 4. Discussions and conslusions \\}

The resulting $m_X$ distributions  for $B\rightarrow R l \nu$ for $R = \pi,
\rho, D, D^{*}, D^{**}$  are shown in Fig.~1, along with the  inclusive
$m_X$  distribution for the  $b \rightarrow u$ transition,
with $| V_{ub} / V_{cb} | = 1$.
The $b \rightarrow c$ transition is
dominated by the $X_{c} = D, D^{*}, D^{**}$,   and can be  reliably
calculated in the HQET as described in the previous section.
The regions between the  triangles are the range of the predicted
rate when the $dm_X$ integration over the delta function is performed.
On the other hand,  the $b \rightarrow u$ transition is largely nonresonant.
The cases with $X_{u} = \pi, \rho$ are shown explicitly for two different
models discussed in Section 3.   For $X_{u} = \pi$, the region between
the open triangles are predictions by Hagiwara {\it et al.} \cite{hagiwara},
and the region   between the closed   triangles are  predictions by
Casalbuoni {\it et al.} \cite{casalbuoni}.
For $X_{u} = \rho$, the regions between the lower two curves are
predictions by Hagiwara {\it et al.}, whereas the region between
the upper  two curves are predictions   by Casalbuoni {\it et al.}.
The inclusive $m_{X}$ distribution for $b \rightarrow u$ was obtained
from the ACCMM model with hadronic mass constraint of
$m_X \stackrel{>}{\sim} 2 m_{\pi}$.
The excusive decay of $B\rightarrow \pi l \nu$ is  shown separately.

{}From Fig.~1, most of the $b\rightarrow u$ transition events survive the cut
on the hadronic invariant mass, $m_{X} < m_{D}$, contrary to the more
conventional cut  on the electron energy.
In fact, one can relax the condition   $m_{X} < m_{D}$ because the
$m_X$ distribution in $b \rightarrow c l \nu$ is completely dominated by
contributions by three resonances $D, D^{*} $ and $D^{**}$, which are
essentially like delta functions, Eq.~(2).
In other words, one can use the $b \rightarrow u$ events in the region
even above $m_{X} = m_D$, excluding small regions in $m_X$ around $m_{X} =
m_{D}, m_{D^{*}}, m_{D^{**}}$.  The cut on the $m_X$ is more
effective  than the cut on the electron energy by factor of $\sim 10$,
and therefore we have much better statistics.
Furthermore, theoretical  understanding of exclusive decay modes of $B
\rightarrow X_{u} l \nu$ is rather poor, as we discussed in Section 3.
Two different models lead to vastly different predictions for $X_{u} =
\pi$ and $\rho$.  This  would induce theoretical uncertainties in
determination of $V_{ub}$ from the measurement of an exclusive
semileptonic decay of $B$ mesons.
On the  other hand, the inclusive decay is better understood, so it
would be more reliable to calculate the $m_X$ distribution for inclusive
$b\rightarrow u$ transitions.

At future $B-$factory with microvertex detectors (symmetrical or
asymmetrical),   one expects that the
efficiency for the event reconstruction will be improved (might 30 \%
efficiency).  Then, among $10^8~~ B \bar{B}$ events,  $\sim  10^5$
events without any constraint on  $m_{X}$ may be reconstructed.
For more details on the problems  of experimental reconstruction and  continuum
background, see Ref.  \cite{cskimetal}.

Even without a full event reconstruction, one may have good measurements
of missing energy and momentum of missing neutrino, $E_{\nu}$ and
$\vec{p}_{\nu}$ satisfying the mass--shell condition,
\[
E_{\nu}^{2} - \vec{p}_{\nu}^{2} = m_{\nu}^{2} = 0.
\]
In this case, the hadronic mass $m_X$ is not fully constructed, but
it is bounded by
\begin{equation}
m_{X}^{2}  <  m_{X,max}^{2} = m_{B}^{2} + m_{l\nu}^{2} - 2 \gamma m_{B}
(E_{l\nu} - \beta p_{l\nu}).
\end{equation}
Here, $m_{B}$ is the $B-$meson mass,  and $\gamma = (1-\beta^{2})^{-1/2}
= m_{\Upsilon} / 2 m_{B}$ for the symmetric $B$ factory with
$e^{+} e^{-} \rightarrow \Upsilon (4S) \rightarrow B \bar{B}$.
Since $\beta$ is very small, $m_{X,max}^2$ is close to $m_{X}^2$, and we
lose very little efficiency.  It turns out that $\sim 80 \%$ events
for $b\rightarrow u$ transitions survive  the cut on the $m_{X,max}^2$
\cite{cskimetal} :
\[
m_{X,max} < m_D.
\]

In any case,  studying the hadronic mass distributions in inclusive
semileptonic $ b \rightarrow u $ transition is  experimentally viable.
It's also theoretically better described, so  theoretical uncertainties
in determining  $|V_{ub}|$  would be less compared to the $|V_{ub}|$
determined from studies of  exclusive  decay modes.
In summary,  we would have  better statistics in extracting $|V_{ub}|$
by measuring the $m_X$ distributions in inclusive $b\rightarrow u$
semileptonic decays,   
and have better theoretical
handles over the inclusive decays rather than exclusive decays.

\vspace{.8in}

\noindent
{\Large \bf Acknowledgements \\}

\noindent
We would like to thank A.I. Vainshtein for  helpful discussions.
This work  was supported
in part by the Korean Science and Engineering  Foundation,
in part by Non-Direct-Research-Fund, Korea Research Foundation 1993,
in part by the Center for Theoretical Physics, Seoul National University,  and
in part by the Basic Science Research Institute Program, Ministry of Education,
  1994,  Project No. BSRI-94-2425.

\vspace{.8in}

\noindent
{\Large \bf Figure Captions}

\vspace{.3in}

{\bf Fig.~1} The $m_X$ distributions  in $B \rightarrow X_{c,u} l \nu$
with $| V_{ub} / V_{cb} | = 1$.  The $b \rightarrow c$ transition is
dominated by the $X_{c} = D, D^{*}, D^{**}$,   and can be  reliably
calculated in the HQET.  The regions between the arrows are predicted
rate in the unit of $10^{13}~{\rm sec}^{-1}$
when the $dm_X$ integration over the delta function is performed.
On the other hand,  the $b \rightarrow u$ transition is largely nonresonant.
The cases with $X_{u} = \pi, \rho$ are shown explicitly for two different
models.   For $X_{u} = \pi$, the region between the open triangles are
predictions by Hagiwara {\it et al.}, and the region   between the closed
triangles are  predictions by Casalbuoni {\it et al.}.  For $X_{u} = \rho$,
the regions between the lower two curves are predictions by Hagiwara {\it et
al.}, whereas the region between the upper two curves are predictions
by Casalbuoni {\it et al.}.  The inclusive $m_{X}$ distribution for
$b \rightarrow u$ was obtained from the ACCMM model with hadronic mass
constraint of $m_X \stackrel{>}{\sim} 2 m_{\pi}$.

\end{document}